\documentclass[preprint,superscriptaddress,amsmath,amssymb,longbibliography,floatfix]{revtex4-2}
\usepackage{graphicx}% Include figure files
\usepackage{bm}% bold math
\usepackage[running]{lineno}
\usepackage{longtable} %Inserted by SC on 22/10/2019
\usepackage{multirow} %Inserted by SC on 22/10/2019
\usepackage{amsmath} %Inserted by SC on 24/10/2019
\usepackage[T1]{fontenc} %Inserted by SC on 24/10/2019
\usepackage{natbib} %Inserted by SC on 02/11/2019
\usepackage{float} %Inserted by SC on 04/11/2019
\usepackage{xcolor} %Inserted by SC on 15/11/2019
\usepackage{relsize} %Inserted by SC on 15/11/2019
\usepackage{url} %Inserted by SC on 18/11/2019
\usepackage[hidelinks]{hyperref}
\hypersetup{
    colorlinks,
    linkcolor={blue},
    citecolor={blue},
    urlcolor={blue}
}
\begin{document}

%\linenumbers

\title{Quantifying small-scale anisotropy in turbulent flows}% Force line breaks with \\

\author{Subharthi Chowdhuri}
    \email{subharc@uci.edu}
    \affiliation{Department of Civil and Environmental Engineering, University of California, Irvine, CA 92697, USA}

\author{Tirtha Banerjee}
    \affiliation{Department of Civil and Environmental Engineering, University of California, Irvine, CA 92697, USA}

\date{\today}% It is always \today, today,

\begin{abstract}
The verification of whether small-scale turbulence is isotropic remains a grand challenge. The difficulty arises because the presence of small-scale anisotropy is tied to the dissipation tensor, whose components require the full three-dimensional information of the flow field in both high spatial and temporal resolution, a condition rarely satisfied in turbulence experiments, especially during field scale measurement of atmospheric turbulence. To circumvent this issue, an \emph{intermittency-anisotropy} framework is proposed through which we successfully extract the features of small-scale anisotropy from single-point measurements of turbulent time series by exploiting the properties of small-scale intermittency. Specifically, this framework quantifies anisotropy by studying the contrasting effects of burst-like activities on the scale-wise production of turbulence kinetic energy between the horizontal and vertical directions. The veracity of this approach is tested by applying it over a range of datasets covering an unprecedented range in the Reynolds numbers ($Re \approx 10^{3}$ to $10^{6}$), sampling frequencies (10 kHz to 10 Hz), surface conditions (aerodynamically smooth surfaces to typical grasslands to forest canopies), and flow types (channel flows, boundary layer flows, atmospheric flows, and flows over forest canopies). For these diverse datasets, the findings indicate that the effects of small-scale anisotropy persists up to the integral scales of the streamwise velocity fluctuations and there exists a universal relationship to predict this anisotropy from the two-component state of the Reynolds stress tensor. This relationship is important towards the development of next-generation closure models of wall-turbulence by incorporating the effects of anisotropy at smaller scales of the flow.
\end{abstract}

\maketitle

\section{Introduction}
\label{Intro}
According to Kolmogorov's hypothesis, the small-scale (comparable to inertial subrange and dissipative scales) turbulence statistics are isotropic, independent of the large-scale (comparable to the integral scales) conditions, and possess universal characteristics \citep{tennekes1972first,renner2002universality}. In this context, isotropy implies that the turbulence statistics are independent of direction, and therefore, should not be affected if the coordinate system is rotated or translated \citep{taylor1935statistical}. The expectation of isotropy at smaller scales of the flow stems from the physical consideration that due to the cascading process the flow at smaller scales loses the memory of anisotropy that persists at the largest scales of the flow \citep{tennekes1972first}. Although the assumption of small-scale isotropy is the backbone of turbulence research, there currently exists no consensus on how to verify whether the small-scales are isotropic or not \citep{biferale2005anisotropy}.

Some previous studies attempted to study the problem of small-scale isotropy through the spectral or structure function methods. In this approach, a scalewise description of turbulence is obtained and the local isotropy hypothesis in the inertial wavenumber range is investigated by employing a few standard measures, such as: studying the $4/3$ ratio of spectral amplitudes; existence of the Kolmogorov $-5/3$ or $+2/3$ power laws in the spectra or second-order structure functions; the rolling off of the momentum cospectra faster than the energy spectra; the Kolmogorov $4/5$ law in the third-order structure functions, and so on \citep{saddoughi1994local,chamecki2004local}. However, none of these measures conclusively show the evidence of local isotropy at small scales, since the inferences obtained from one measure differ from the other \citep{chamecki2004local}.

Apart from these measures, an another popular approach has been to consider a statistical quantity named dissipation tensor, whose properties are quite sensitive to the presence of small-scale eddies. The anisotropic dissipation tensor ($d_{ij}$) is defined in a Cartesian co-ordinate system as, 
\begin{align}
\begin{split}
d_{ij} =\frac{\epsilon_{ij}}{\epsilon_{ii}}-\frac{1}{3}\delta_{ij}, \ \epsilon_{ij}=2\nu\overline{\frac{\partial{{u^{\prime}}_{i}}}{\partial{x_{k}}}\frac{\partial{{u^{\prime}}_{j}}}{\partial{x_{k}}}},
\end{split}
\label{tensor}
\end{align}
where overbar indicates averaging over time or space, $u^{\prime}_{i}$ are the turbulent fluctuations in the velocity field ($i=1,2,3$), $\delta_{ij}$ is the Kronecker delta, $\epsilon_{ij}$ is the dissipation rate of $\overline{{u^{\prime}}_{i}{u^{\prime}}_{j}}$, and $\nu$ is the kinematic viscosity of the fluid. This tensor becomes zero in isotropic turbulence and its anisotropy is quantified by using the invariants of $d_{ij}$, an approach known as invariant analysis \citep{gerolymos2016dissipation}. However, except a few studies \citep{gulitski2007velocity,longo2017invariants}, the measurements of all the nine components of the dissipation tensor are practically very hard to obtain from point-based observations. Therefore, these are mainly evaluated from the direct numerical simulations of turbulent flows \citep{smyth2000anisotropy,liu2008anisotropy,djenidi2012anisotropy}.

Given the problems with the estimation of dissipation tensor and the uncertainties associated with other measures, an alternative assessment of small-scale anisotropy is sought whose foundations are rooted in the phenomenology of small-scale turbulence. One such aspect of anisotropy is related to small-scale intermittency, characterized by the appearance of strong non-Gaussian tails in the velocity increments as the scales of the flow tend to decrease \citep{sreenivasan1997phenomenology,buaria2022intermittency}. The presence of small-scale intermittency introduces anomalous scalings in the structure function moments, rendering them to be significantly different from the ones arising from the assumption of local isotropy at smaller scales of the flow \citep{sreenivasan1997phenomenology,tordella2011small}. 

\citet{carter2017scale} exploited the concept of intermittency to study small-scale anisotropy in homogeneous turbulence. By analysing the particle image velocimetry (PIV) measurements in three different directions (i.e., streamwise, spanwise, and vertical), \citet{carter2017scale} studied the effects of intermittency on the higher-order structure functions of the velocity components. The authors found that the effect of intermittency was quite sensitive to the direction being considered, and therefore, linked such behaviour with the presence of small-scale anisotropy. Although this study explored an alternate way to characterize small-scale anisotropy, it had certain caveats. First, the study was performed for a homogeneous flow at a very low Reynolds number ($Re \approx 400$). Second, this study assessed anisotropy in a qualitative sense rather than quantifying it through a statistical measure. Third, the authors employed higher-order moments whose estimations require high-resolution measurements, which are not readily available for all flows types, especially for high-Reynolds number atmospheric flows. 

On the other hand, in this current work, we extend the concept of \emph{intermittency-anisotropy} to inhomogenous wall-turbulence, where the turbulence statistics are known to depend on the wall-normal locations and a directional bias exists between the horizontal and vertical directions \citep{smits2011high}. For such flows, we specifically show that the small-scale anisotropy can be comprehensively studied by only considering the directional effects of burst-like activities on the turbulence kinetic energy at each scale of the flow. Therefore, we limit ourselves to the second-order moments, whose computations do not necessarily require high-resolution measurements. To test the robustness of our approach, a large corpus of experimental and numerical datasets from wall-turbulence are used, covering an unprecedented range in the Reynolds numbers ($10^{3}$ to $10^{6}$), sampling frequencies (10 kHz to 10 Hz), surface conditions (aerodynamically smooth surfaces to typical grasslands to forest canopies), and flow categories (channel flows, boundary layer flows, atmospheric flows, and flows over forest canopies). By analyzing these diverse datasets (see Table \ref{tab:1} for a summary), our objectives are primarily threefold. First, to investigate up to what scales the small-scale effects persist in turbulent flows and whether that scale is universal. Second, to formulate a bulk measure of small-scale anisotropy for these wide class of flows by exploiting the phenomenology of small-scale turbulence. Third, to propose a diagnostic relationship to predict small-scale anisotropy from the large-scale conditions itself. To achieve these goals, this study is organized in three different sections. In Section \ref{data_method} we introduce the different datasets and our framework. In Section \ref{results}, the results are presented and discussed to elucidate on the flow physics. Finally, in Section \ref{conclusion} the conclusions and scopes for future research are outlined. 

\section{Dataset and methodology}
\label{data_method}
\subsection{Dataset}
\label{Data}
\subsubsection{Channel and boundary-layer flows}
\label{Eng_flows}
To accomplish our objectives, we use two datasets from a turbulent channel flow and from a boundary-layer flow. The first of these datasets is a numerical one, obtained from direct numerical simulation (DNS). The simulation was carried out at a Reynolds number of $Re= 2003$ and the resulting dataset is available at \citep{hoyas2006scaling}. The simulation is run on a smooth-wall channel setup with periodic boundary conditions in the streamwise ($x$) and spanwise ($y$) directions. The domain size of the simulation is $8\pi \delta \times 2\delta \times 3\pi \delta$ in the streamwise, vertical, and spanwise directions respectively, where $\delta$ is the half-channel height. The numerical grid consists of $6144$ and $4608$ uniformly-spaced grid points in the streamwise and spanwise direction, respectively, while a non-uniform grid with $633$ points is used in the wall-normal direction ($z$). We refer to this dataset as the DNS dataset and carry out our computations on the streamwise direction and average the results over multiple spanwise locations. This strategy is adopted to ensure that the results obtained from DNS can be directly compared with other point flow set-ups. 

The other dataset is an experimental one from a fully-developed turbulent boundary layer flow over an aerodynamically smooth flat plate, as obtained in the wind-tunnel facility of the University of Melbourne \citep{MARUSIC2020dataset}. The Reynolds number of this flow is of the order of $Re\approx 10^4$. Regarding this experiment, only the time series of the streamwise velocity, $u$, are available from hot-wire measurements at a sampling frequency of $20$\,kHz for up to $120$-s. Further details of the experiment can be found in \citet{baars2015wavelet} and we refer to this dataset as the TBL dataset. For both DNS and TBL datasets, the turbulent fluctuations ($x^{\prime}$, where $x=u,v,w$) are computed by subtracting the spatial (time)-averaged mean velocity from $x$. Throughout this study, the wall-unit normalization is indicated by the $+$ superscript such that $z^+=zu_{*}/\nu$, where $u_{*}$ is the friction velocity and $\nu$ is the kinematic viscosity of air.

\subsubsection{Atmospheric surface layer flows}
\label{atm_flows}
Regarding atmospheric flows, we use five different datasets from the meteorological masts positioned within the surface layer. These flows are categorized as ASL flows, and by assuming the depth of the atmospheric boundary layer to be around 500 m, the Reynolds numbers of these flows roughly correspond to $Re \approx 10^{6}$. One of these datasets is collected during the SLTEST experiment, where nine North-facing time-synchronized CSAT3 sonic anemometers were mounted on a 30-m mast, spaced logarithmically over an 18-fold range of heights, from 1.42 m to 25.7 m, with the sampling frequency being set at 20 Hz \citep{chowdhuri2019representation,chowdhuri2023revisiting}. The other dataset is from a field experiment conducted over Oceano Dunes in California, where sonic anemometer observations were available from a 10-m tower with a sampling frequency of 50 Hz \citep{martin2017wind,martin2018high,huang2019fine}. Both of these measurement sites were topographically flat and aerodynamically smooth.  We refer to these datasets as SLTEST and Oceano, respectively. The other two experimental datasets were obtained during an experimental campaign (CAIPEEX-IGOC) in India, and high-frequency observations of the three velocity components were collected at a sampling frequency of 10-Hz \citep{chowdhuri2019evaluation}. The site conditions were representative of a typical grassland and we refer to these experiments as CPX1 and CPX2, respectively. One another experimental dataset is used, collected over a grassland at the Blackwood division of the Duke Forest in Durham, North Carolina with a sampling frequency of 56 Hz \citep{katul1997energy}. This dataset is simply referred to as Grass. For our purposes, we restrict all these observations to near-neutral conditions, i.e., when the effect of buoyancy is negligible. It is done to ensure that the ASL results can be compared effectively with the channel and boundary-layer flows. The results reported in Section \ref{results} are averaged over an ensemble of near-neutral runs of 30-min duration each. The near-neutral runs are identified as those satisfying the condition $|z/L|<0.5$, where $L$ is the Obukhov length and $z$ is the observation height. 

\subsubsection{Roughness sublayer flows}
\label{RSL_flows}
In order to account for the effect of roughness, we use three different datasets where measurements were carried out within the roughness sublayers. These flows are labelled together as RSL flows and their Reynolds numbers too are of the order of $10^6$. One of these datasets is the GoAmazon one, where nine level measurements were available over a dense Amazon forest \citep{fuentes2016linking,freire2017turbulent,gerken2018investigating,chowdhuri2022scale}. The measurement heights are within the range of $z/h=0.2-1.38$, where $h$ is the height of the trees, approximately equal to 35 m. The leaf area index (LAI), which is defined as the total one-sided leaf area (half the total foliage area) per unit ground surface area, is estimated to be between 6.1 and 7.3 m$^{2}$ m$^{-2}$. The other dataset is over a maize canopy, where five observation heights are available at $z/h=1/3,2/3,3/3,4/3,5/3$, with $h$ being equal to 2.05 m \citep{chamecki2013persistence,gleicher2014interpreting}. The LAI for the maize canopy is around 3.3 m$^{2}$ m$^{-2}$ \citep{chamecki2013persistence}. For both GoAmazon and maize canopies, the sampling frequencies of the measurements are set at 25 Hz. A third dataset is over Loblolly pine canopies in Duke forest, where only one measurement height is available at $z/h=1.44$, where $h$ is the height of the pine trees (13 m) and the sampling frequency is set at 10 Hz \citep{katul1997turbulent}. We refer to this dataset as DF and the LAI for this forest is 3.1 m$^{2}$ m$^{-2}$. These various LAI values indicate how different the canopy structure was among the GoAmazon, Maize, and DF datasets. For all these three RSL datasets, we restrict ourselves to near-neutral stratification (satisfying $|(z-d)/L|<0.5$, where $d=2h/3$) and perform an average over an ensemble of 30-min runs belonging to such conditions. For convenience, all these diverse datasets are summarized in Table \ref{tab:1}. 

\begin{table}
\begin{center}
\begin{tabular}{|c |c | c |c |c |c|}
\hline
Dataset  & Flow type  & Surface & Variables & $f_s$ & Ensemble\\
\hline
DNS \citep{hoyas2006scaling} & Channel & Smooth & $u$, $v$, $w$ & 256 Hz & NA \\
\hline
TBL \citep{MARUSIC2020dataset} & Boundary-layer & Smooth & $u$ & 10 kHz & NA \\
\hline
SLTEST \citep{chowdhuri2023revisiting} & ASL & Saltland & $u$, $v$, $w$ & 20 Hz & 20\\
\hline
Oceano \citep{huang2019fine} & ASL & Sand & $u$, $v$, $w$ & 50 Hz & 611 \\
\hline
CPX1 \citep{chowdhuri2019evaluation} & ASL & Grassland & $u$, $v$, $w$ & 10 Hz & 130 \\
\hline
CPX2 \citep{chowdhuri2019evaluation} & ASL & Grassland & $u$, $v$, $w$ & 10 Hz & 160 \\
\hline
Grass \citep{katul1997energy} & ASL & Grassland & $u$, $v$, $w$ & 56 Hz & 100 \\
\hline
GoAmazon \citep{chowdhuri2022scale} & RSL & Forest ($h=$ 35 m) & $u$, $v$, $w$ & 20 Hz & 93 \\
\hline
Maize \citep{chamecki2013persistence} & RSL & Crop ($h=$ 2.05 m) & $u$, $v$, $w$ & 20 Hz & 16 \\
\hline
DF \citep{katul1997turbulent} & RSL & Forest ($h=$ 13 m) & $u$, $v$, $w$ & 10 Hz & 160 \\
\hline
\end{tabular}
\caption{A summary of different datasets used in this study. The variables $u$, $v$, and $w$ denote the velocity components in streamwise, spanwise, and vertical directions, respectively. Here $f_s$ indicates the sampling frequencies, and for DNS this quantity refers to the inverse of the streamwise spacing. Ensemble specifies the number of 30-min near-neutral runs being used to average the results and are only applicable for ASL and RSL flows. The symbol $h$ denotes the canopy height.}
\label{tab:1}
\end{center}
\end{table}

\subsection{Methodology}
\label{method}
To quantify small-scale intermittency, we adopt the burst framework introduced by \citet{chowdhuri2023revisiting}. Since this framework has already been discussed in detail, we briefly summarize the important concepts here. For a velocity signal $u^{\prime}$, the effects of strong fluctuations on its instantaneous energy content is quantified by drawing a plot between the cumulative distributions of duration against its amplitudes. In this context, duration ($t_{p}$) is simply defined as those time instances up to which the signal stays positive or negative. The amplitudes ($S^{2}_{p}$), on the other hand, are defined as,
\begin{equation}
S^{2}_{p}=\frac{1}{T \times \overline{ {u^{\prime}(t)}^2}}{\int_{t}^{t+t_{p}} {u^{\prime}}^2(t) \,dt},   
\end{equation}
where $T$ is the total duration of the time-series. The quantity $S^{2}_{p}$ represents the contribution from an event of duration $t_p$ to the instantaneous variance ${u^{\prime}}^2(t)$.

When the cumulative distributions of $t_p$ and $S^{2}_{p}$ are plotted against each other, if no amplitude effect was present they would follow a straight line with $45^{\circ}$ slope. Therefore, the departure from this straight line statistically represents the effect of strong bursts in the flow and is quantified as an area between the curve and the straight line. This area is termed as burstiness index ($\mathcal{B}^2_{u}$) and is bounded between 0 to 0.5. The concept of burstiness index can be extended to create a scale-wise description ($\mathcal{B}^2_{\Delta u} (\tau)$), where the burstiness curves are constructed for the signals $\Delta u=u^{\prime}(t+\tau)-u^{\prime}(t)$, which represents the velocity increments at a scale $\tau$. Physically, $\mathcal{B}^2_{\Delta u} (\tau)$ represents the effect of bursts to the turbulence kinetic energy at each scale of the flow.

\begin{figure*}[h]
\centerline{\includegraphics[width=\textwidth]{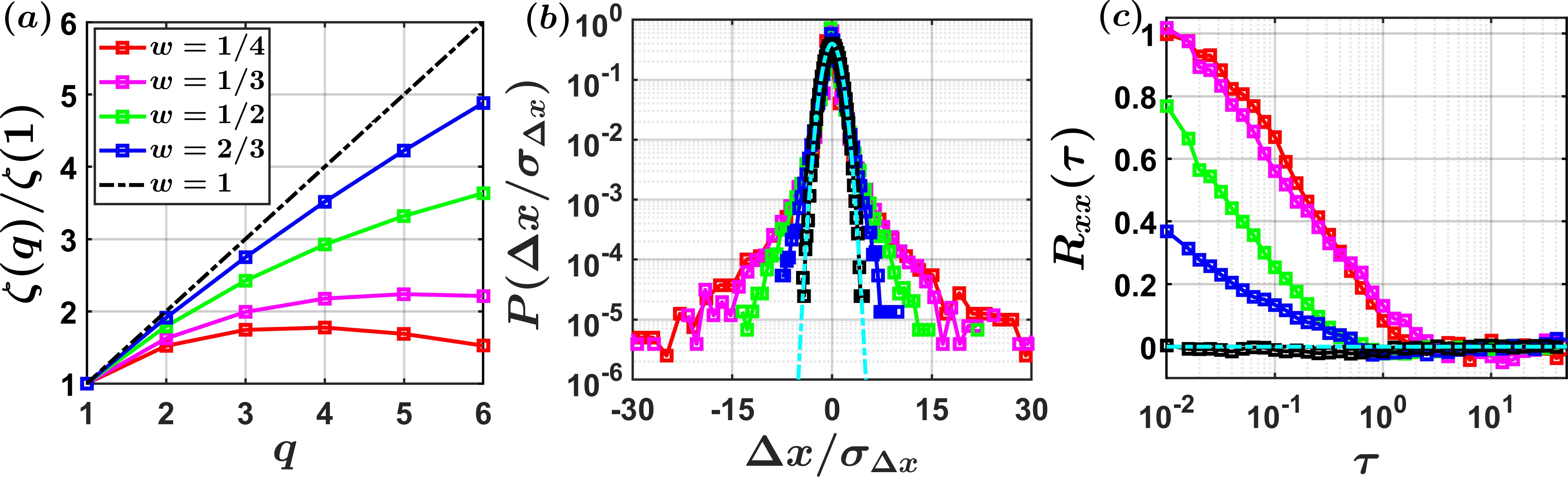}}
\caption{The plots of (a) structure function moments ($\zeta(q)$, where $q$ are the moment orders), (b) the PDFs of the signal increments ($P(\Delta x/\sigma_{\Delta x})$, where $\Delta x$ are the increments and $\sigma$ denote their standard deviations) and (c) the scale-wise variations in small-scale intermittency ($R_{xx}(\tau)$, where $\tau$ are the time lags) are shown for such synthetic signals whose level of intermittency is controlled by changing a tunable parameter $w$. The legend in (a) denotes the different $w$ values. The cyan colored dash-dotted line in (b) indicates the Gaussian distribution.}
\label{fig:1}
\end{figure*}

This concept can now be modified to generate a scale-wise description of small-scale intermittency. Small-scale intermittency is characterized by non-Gaussian distributions of the velocity increments \citep{sreenivasan1997phenomenology}. This is illustrated through Fig. S1 in \citep{Note34}, where for all the previously defined datasets the velocity increments of streamwise ($\Delta u$) and vertical velocity ($\Delta w$) fluctuations display strong non-Gaussian tails. However, this non-Gaussian distribution can be destroyed through a Fourier phase randomization operation, by converting them to a Gaussian distribution \citep{poggi2009flume}. One can employ an iteratively adjusted amplitude Fourier transform (IAAFT) model for this purpose, which preserves the probability density functions (PDFs) of the signal and its spectrum but destroys the effects of large non-Gaussian tails in the velocity increments \citep{basu2007estimating}. It is, therefore, tempting to consider a ratio of the burstiness indices between the original and IAAFT signal at each scale and if this ratio deviates from unity that would be solely due to the presence of small-scale intermittency.

To test this hypothesis, we generate a family of synthetic turbulent signals (with each consisting of $10^6$ data-points) from a log-Poisson cascade model, which has a tunable parameter $w$ and by changing it systematically one can either increase or decrease the effects of small-scale intermittency. The details of this model can be found in \citet{chainais2003scale}. One of the tell-tale signs of intermittency is the departure from the Kolmogorov prediction of the structure function moments ($\zeta(q)$, where $q$ are the moment orders). From Fig. \ref{fig:1}a, one can clearly see that by decreasing $w$, large departures from the Kolmogorov prediction ($\zeta(q)/\zeta(1)=q$, where $\zeta(q)=q/3$) is observed, which is represented by a straight line of $45^{\circ}$ slope. According to \citet{chainais2003scale}, the moments $\zeta(q)$ for different $w$ values of the log-Poisson cascade model can be written as, 
\begin{equation}
    \zeta(q)=(2w-1)\frac{q}{3}+2(1-w^{q/3}),
\end{equation}
which are shown in Fig. \ref{fig:1}a as colored lines after normalizing them with $\zeta(1)$. Moreover, if the PDFs of the normalized signal increments ($\Delta x/\sigma_{\Delta x}$, where $\sigma_{\Delta x}$ is the standard deviation) are plotted, the lowest $w$ values correspond to the heaviest tails while the ones with $w=1$ follow a Gaussian distribution (Fig. \ref{fig:1}b).

Based on our hypothesis, we compute the ratios of the burstiness indices at each scale $\tau$, defined as,
\begin{equation}
    R_{xx}(\tau)=\frac{\mathcal{B}^2_{\Delta x}}{\mathcal{B}^2_{\Delta x_{\rm p}}}-1,
    \label{R_curve}
\end{equation}
where the subscript $\rm p$ indicates the burstiness index of an IAAFT signal. Note that we subtract 1 so that the ratios are zero when no difference exists between the original and IAAFT signals. As expected, $R_{xx}(\tau)$ deviates the strongest from 0 for the lowest $w$ values (Fig. \ref{fig:1}c). On the other hand, $R_{xx}(\tau)$ systematically decrease with increasing $w$, and eventually for $w=1$, they remain at zero irrespective of the scales under consideration. Hence, $R_{xx}(\tau)$ is indeed sensitive to the presence of heavier tails in the distributions of signal increments and can be used to generate a scale-wise description of small-scale intermittency. The results related to that aspect are presented for real turbulent signals in Section \ref{results}.

\section{Results and discussion}
\label{results}
We begin with discussing up to what scale does the effect of small-scale intermittency persist. For this purpose, we use a scale-wise description of intermittency based on the framework described earlier. This information is further utilized to construct a metric for small-scale anisotropy, and a diagnostic relationship is proposed to estimate this quantity from the anisotropic states of the Reynolds stress tensor for a wide range of wall-bounded flows.

\subsection{Scale-wise description of small-scale intermittency}
\label{Res1}
\begin{figure*}
\centerline{\includegraphics[width=\textwidth]{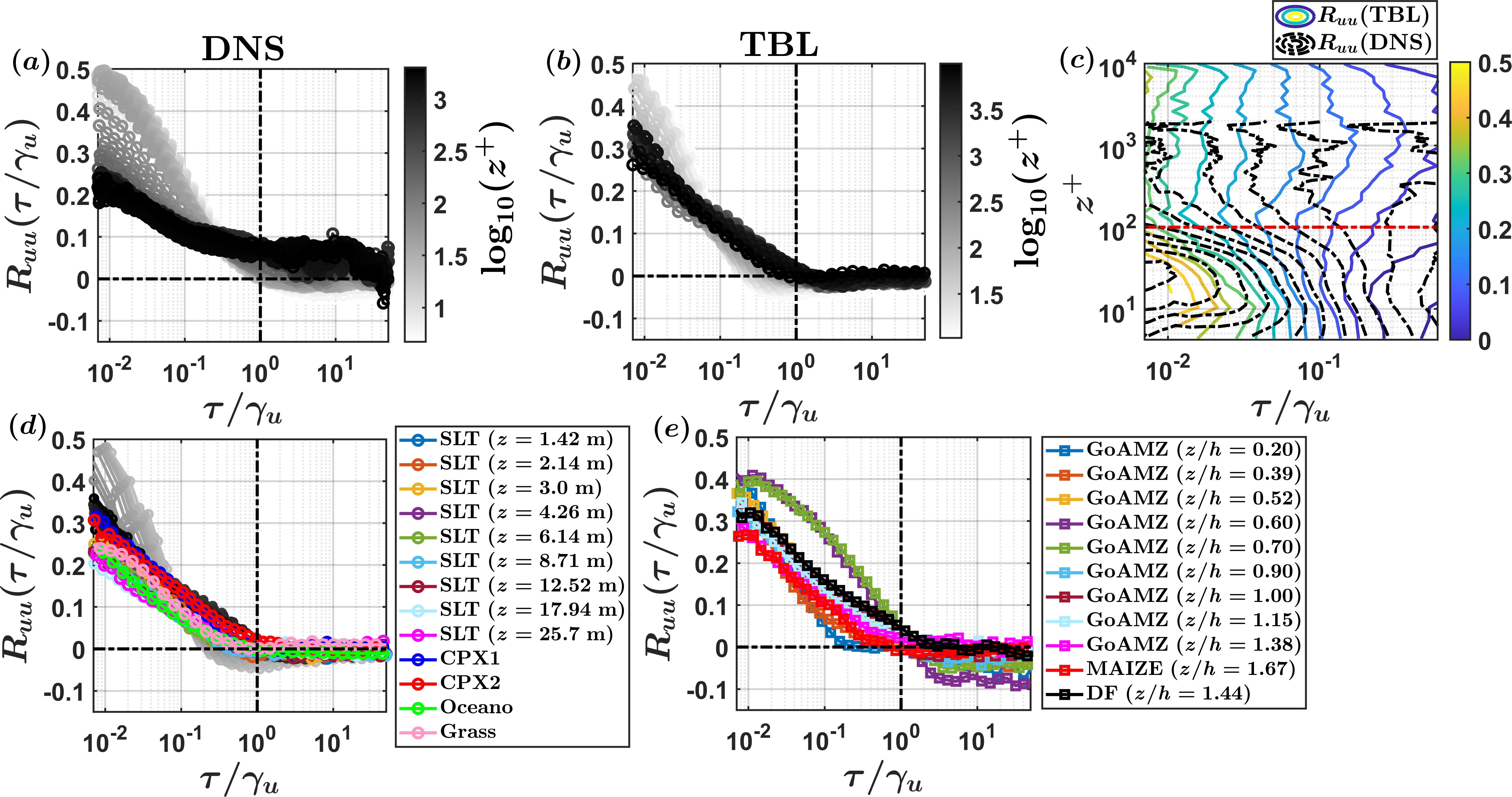}}
\caption{The scale-wise intermittency curves of streamwise velocity fluctuations, as quantified through $R_{uu}$, are shown for the (a) DNS and (b) TBL datasets. The time scales ($\tau$) are normalized by the integral scales of $u^{\prime}$ ($\gamma_{u}$). The grey color-bars represent the logarithms of the wall-normal heights, $\log_{10}(z^{+})$, with the intensities increasing as $z^{+}$ increases. (c) The intermittency curves are compared between the DNS and TBL datasets through a contour plot, where the contours denote the $R_{uu}$ values, and the $x$ and $y$ axes represent $\tau/\gamma_{u}$ and $z^{+}$, respectively. The colored contours correspond to the TBL dataset while the black ones are for the DNS dataset. (d) The $R_{uu}$ curves from different ASL datasets are overlaid on the TBL dataset, indicated through different colors as shown in the legend. (e) The $R_{uu}$ curves corresponding to different canopy datasets (see the legend) are shown, belonging to the RSL flow category. The $h$ indicates the heights of the canopies.}  
\label{fig:2}
\end{figure*}

To assess the scales ($\tau$) up to which the small-scale effects continue, we first compare the $R_{uu}$ curves between the DNS and TBL datasets. These $R_{uu}(\tau)$ values are obtained similarly as in Eq. \ref{R_curve} but for the $u^{\prime}$ signals. Regarding DNS dataset, the scales are the spatial ones along the streamwise direction and the curves are averaged over multiple spanwise locations. The integral scales of $u^{\prime}$, computed from their autocorrelation curves, are used as a normalization factor for the time or spatial lags. For both of these datasets, the $R_{uu}$ curves are plotted for heights spanning from the viscous sublayer to the logarithmic layer and are color-coded according to their $\log_{10}(z+)$ values (see the colorbars in Figs. \ref{fig:2}a--b). Specifically, the light gray colors denote the heights within the viscous sublayer, while the ones with more intensities (i.e., the colors approaching black) indicate the heights from the logarithmic layer. 

Upon comparing Figs. \ref{fig:2}a and b, one can clearly see that the $R_{uu}$ values remain the largest within the viscous layers but they decrease systematically as the logarithmic layer is approached. Physically this finding implies that the effects of small-scale intermittency dominate the turbulence statistics the most at the lower layers of the flow. The same conclusion was reached by \citet{onorato2000small}, where they linked this behaviour with the presence of the bursting events in the viscous sublayer. However, for further verification, one could compare the $R_{uu}$ values with the coefficient of variability (COV) of the instantaneous dissipation rate along the streamwise direction, $\varepsilon={({\partial u}/{\partial x})}^2$, since this quantity is sensitive to intermittency at smaller scales of the flow \citep{frisch1995turbulence,jimenez1998small}. Although for DNS data ${({\partial u}/{\partial x})}^2$ can be estimated directly, Taylor's hypothesis is used for the TBL flow to convert the temporal derivative to a spatial one. The results are shown in Fig. S2 of \citep{Note34}, and can be noticed that the strong $R_{uu}$ values correlate nicely with the COV of $\varepsilon$. Therefore, one can confidently claim that the $R_{uu}$ curves indeed encapsulate the effects of small-scale intermittency in turbulent flows. 

However, as opposed to the smaller ones, at larger scales the $R_{uu}$ curves approach zero, irrespective of the wall normal locations. This saturation towards zero is very clear for the TBL data and occurs precisely at scales equal to $\gamma_u$. On the other hand, for the DNS data, in the logarithmic layers, the curves do not saturate exactly at zero since the attainment of the plateau is not very prominent. To investigate this more carefully, in Fig. \ref{fig:2}c, we show the contour plots of $R_{uu}$ values, plotted against $\tau/\gamma_{u}$ and $z^{+}$. To differentiate between the two datasets, the $R_{uu}$ contours of the DNS data are shown in black while for the TBL ones they are color-coded. It can be clearly seen that within the viscous layers (i.e., $z^{+} \leq 100$ as per \citep{wang2021scaling}), the $R_{uu}$ contours agree sufficiently well between the two datasets. But in the logarithmic layers the contours diverge. A possible interpretation of this phenomenon is, as compared to the TBL data, the outer-layer structures in the DNS data are not well developed due to their low Reynolds numbers \citep{iacobello2023coherent}, thereby causing such lack of convergence towards zero in the $R_{uu}$ curves.

After establishing this fact, we next move on to compare the high Reynolds number ASL datasets ($Re \approx 10^6$) with the moderate Reynolds number flow (($Re \approx 10^4$), which is the TBL one. As discussed in Section \ref{Data}, we use several ASL datasets collected over a range of surface conditions, spanning between an aerodynamically smooth surface to a more typical grassland, with the measurement frequencies varying between 10 to 56 Hz. These comparisons are shown in Fig. \ref{fig:2}d. Despite such huge differences in the Reynolds numbers, surface conditions, or sampling frequencies, the $R_{uu}$ curves from several ASL datasets remarkably collapse on to the curves being constructed from the logarithmic layers of the TBL flow. Moreover, in sync with the TBL data, the ASL curves too attain a clear plateau at zero, comparable to scales of the order of $\gamma_u$. Together these observations suggest that, notwithstanding the differences in the large-scale conditions, through our framework one could unravel the universal aspects of small-scale turbulence. This is indeed a significant result since many previous studies found that the traditional statistics of small-scale turbulence (computed through the moments of the structure functions) obtained from the ASL datasets disagree with the low Reynolds number flows \citep{chamecki2004local,freire2023atmospheric}.

An another interesting outcome emerges when one considers the $R_{uu}$ curves from the canopy datasets where the observations are collected over the roughness sublayers, or in other words, known as the RSL flows. In order to restrict the number of lines, we show all the nine observation heights from the GoAmazon data and the only available observation from the Duke Forest ($z/h=1.44$), but for the Maize canopy we only limit ourselves to the observation height of $z/h=5/3$. The entire observations from the Maize canopy are shown in Fig. S3 of \citep{Note34}. From Fig. \ref{fig:2}e, one could see that the $R_{uu}$ curves of the RSL flows are qualitatively similar to the ASL ones, but for two specific heights from the GoAmazon dataset the $R_{uu}$ values remain quite large and thus differ from the rest of the curves. These two heights are at $z/h=$ 0.6 and 0.7, which precisely correspond to the locations where the leaf area densities of the plant elements are the largest \citep{dias2015large}. Therefore, this finding suggests that the eddies created at the wakes of the plant elements (such as leaves, stems, etc.) contribute significantly to the small-scale intermittency of the streamwise velocity components in RSL flows. This conclusion remains true for the Maize canopy as well, where one particular observation height ($z/h=1/3$) stands out from the rest (see Fig. S3a in \citep{Note34}).

\begin{figure*}[h]
\centerline{\includegraphics[width=\textwidth]{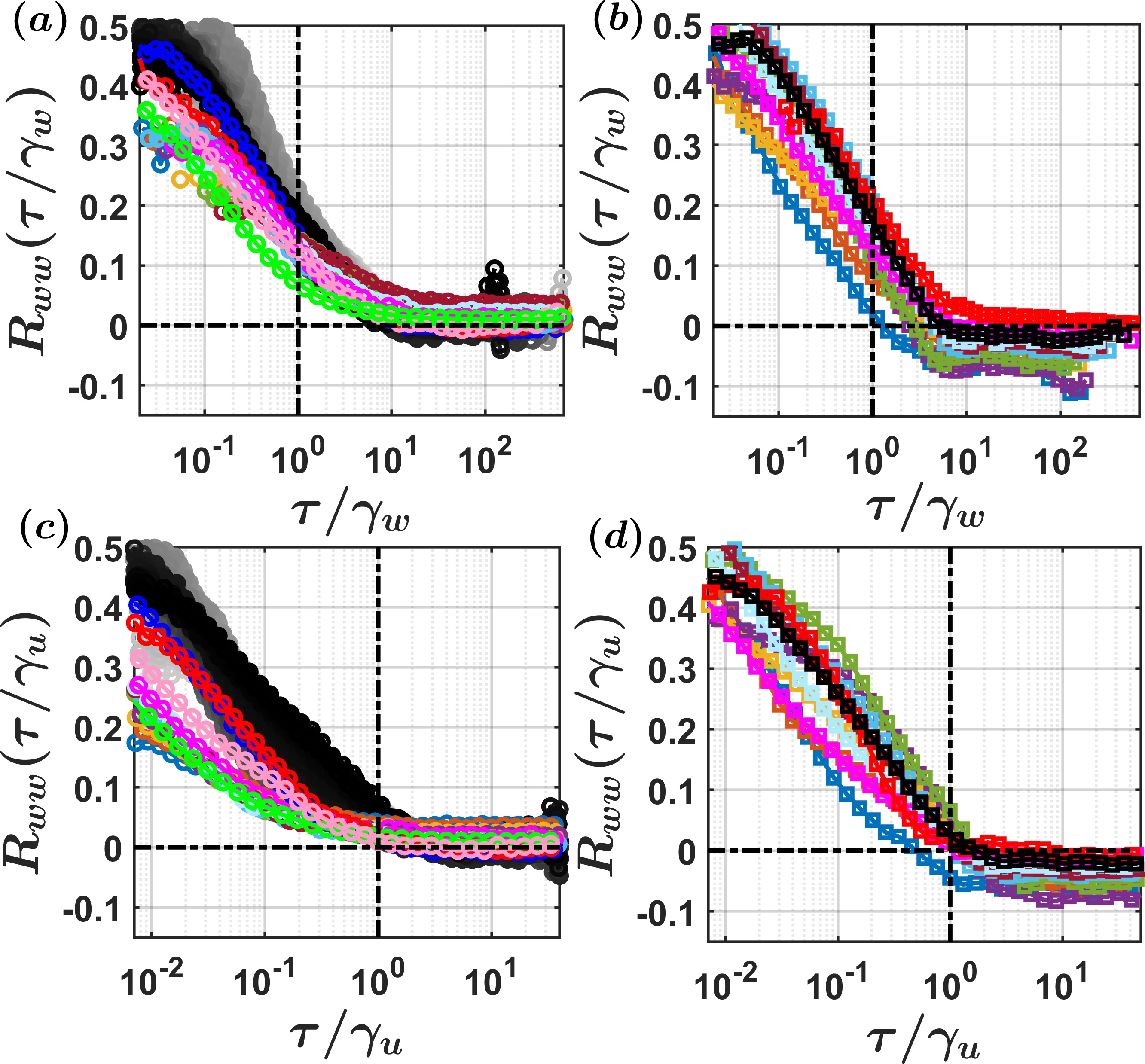}}
\caption{Same as Fig. \ref{fig:2}, but for the vertical velocity fluctuations ($R_{ww}$). (a), (c) The $R_{ww}$ curves are compared between the DNS and ASL datasets, where the grey colored lines represent the DNS dataset while the colored ones are from ASL flows. (b), (d) The $R_{ww}$ curves are shown for different canopy datasets. The timescales in (a)--(b) are normalized by the integral scales of $w$ ($\gamma_{w}$), while $\gamma_{u}$ is used in (c)--(d).}
\label{fig:3}
\end{figure*}

After $R_{uu}$, we now turn our attention towards the $R_{ww}$ curves, whose non-zero values are connected to the effects of small-scale intermittency on the vertical velocity fluctuations ($w^{\prime}$). Since the $w^{\prime}$ measurements were not available for the TBL flow, DNS observations are used for comparing with the ASL datasets. In Fig. \ref{fig:3}a, the $R_{ww}$ curves are shown for both DNS and ASL datasets where the time scales are normalized by $\gamma_{w}$, which is the integral scale of $w^{\prime}$. Regarding DNS, as opposed to $R_{uu}$, no significant differences are observed in the $R_{ww}$ values as one transitions from the viscous sublayer to the logarithmic layer. Furthermore, the $R_{ww}$ curves attain a plateau at 0, while for the $R_{uu}$ curves no such clear indication is evident (see Fig. \ref{fig:2}a). The behaviour of $R_{ww}$ curves remain qualitatively similar between the DNS and ASL datasets, and more importantly, the time scales at which the $R_{ww}$ values saturate towards 0 are significantly larger than $\gamma_{w}$ at least by an order of magnitude. Thus, the effects of small-scale intermittency persists well beyond $\gamma_{w}$. However, by using spectra and second-order structure functions of $w^{\prime}$, a few studies concluded that $\gamma_{w}$ can be used as a scale to separate the inertial-subrange turbulence from the large-scale ones \citep[e.g.,][]{katul1998theoretical,zorzetto2018extremes}. Our results, therefore, put a caution against using this scale to isolate the features of small-scale turbulence. This problem gets resolved when for $R_{ww}$ curves, the time scales are normalized instead by $\gamma_{u}$ and the values attain a plateau at scales comparable to the integral scale of $u^{\prime}$ (Fig. \ref{fig:3}c). 

\begin{figure*}[h]
\centerline{\includegraphics[width=\textwidth]{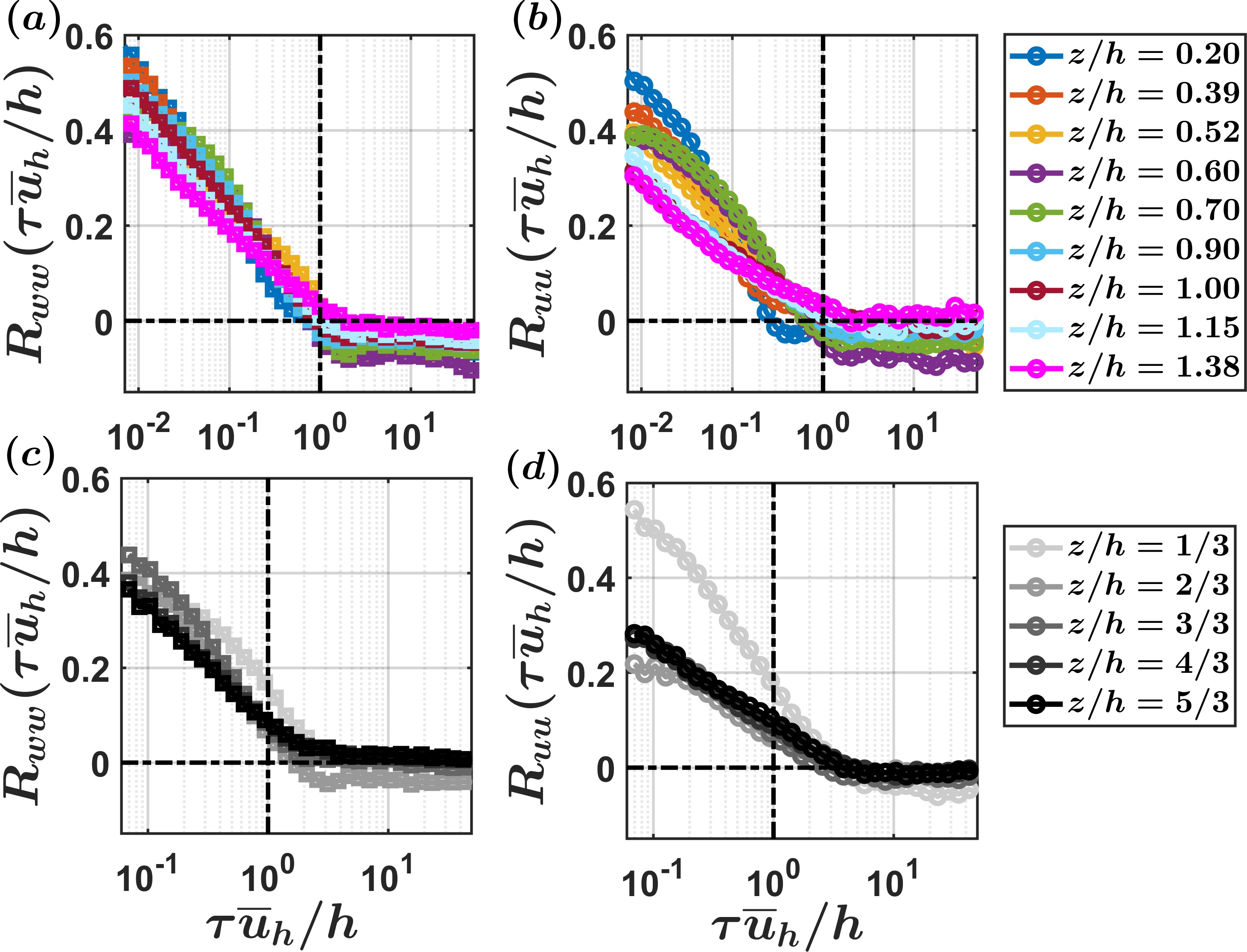}}
\caption{The $R_{ww}$ and $R_{uu}$ curves are shown for the (a, b) GoAmazon and (c, d) Maize canopy datasets. The legends indicate the different heights for both datasets.}
\label{fig:3_1}
\end{figure*}

The same conclusion remains true for the RSL datasets as well (Figs. \ref{fig:3}b and d, and Fig. S3b of \citep{Note34}). Particularly, for these $R_{ww}$ curves, no clear demarcation is observed for $z/h=$ 0.6 and 0.7 levels, as was the case for $R_{uu}$ (Fig. \ref{fig:2}e). In fact, the $R_{ww}$ curves reasonably collapse when the time scales are normalized by the canopy shear scale $h/\overline{u}_{h}$ (where $\overline{u}_{h}$ is the mean wind speed at $z/h=1$), which is the characteristic scale of the canopy-induced coherent structures \citep{raupach1981turbulence,finnigan2000turbulence}. This result is shown in Figs. \ref{fig:3_1}a and c, where for both Maize and GoAmazon canopies similar inferences can be drawn. However, from Figs. \ref{fig:3_1}b and d, it becomes apparent that the canopy shear scale cannot collapse the $R_{uu}$ curves. This suggests that for RSL flows the canopy-scale coherent structures control the intermittency effects for vertical velocity fluctuations, while the eddies created at the wakes of plant elements do the same for $u^{\prime}$. These different scaling properties of $R_{uu}$ and $R_{ww}$ curves are of significant importance since at present no consensus exists on how the presence of canopies influence the small-scale intermittency characteristics. Some studies postulate that the coexistence of canopy-scale coherent structures and small-scale eddies generated at the wakes of plant elements create a complex flow pattern whose intermittency features are different from smooth-wall turbulent flows \citep[e.g.,][]{bolzan2002analysis}. On the other hand, \citet{shnapp2021small} claims that the small-scale intermittency of canopy turbulence shares resemblance with homogeneous and isotropic turbulence. Our results convincingly demonstrate that the presence of a canopy does indeed modify the small-scale features as opposed to the channel, TBL, and ASL flows, albeit differently for the streamwise and vertical velocity fluctuations.

By combing this insight with the differences being observed between $R_{uu}$ and $R_{ww}$ curves for the channel and ASL flows, one could conclude that there exists a directional preference regarding intermittency, such that it has distinct characteristics for the horizontal and vertical velocity fluctuations. This phenomenon is eventually tied to small-scale anisotropy and to quantify that aspect we introduce a concept named intermittent Reynolds stress tensor.

\subsection{Intermittent Reynolds stress tensor}
\label{Res2}
By convention, the anisotropic Reynolds stress tensor quantifies the differences in how the turbulence kinetic energy is distributed among the three coordinate directions. This tensor is defined as, 
\begin{equation}
b_{ij} =\frac{\overline{u^{\prime}_{i}u^{\prime}_{j}}}{2q}-\frac{1}{3}\delta_{ij}, \ q=\frac{\overline{u^{\prime}_{k}u^{\prime}_{k}}}{2}, 
\label{reynolds}  
\end{equation}
where $i=$ 1, 2, and 3 denote the streamwise, cross-stream, and vertical directions, $q$ is the turbulent kinetic energy, and $\delta_{ij}$ is the Kronecker delta. To quantify anisotropy, the eigen values of $b_{ij}$ are used and any difference between the three eigenvalues is connected to the anisotropy of the velocity field. This anisotropy measure is sensitive to the large-scale flow features, such as the presence of coherent structures \citep{dey2018turbulent,konozsy2019new}. Analogously, for our purposes, we define an intermittent Reynolds stress tensor ($\tilde{b_{ij}}(\tau)$) whose components are more sensitive to the presence of small-scale intermittency. This new tensor is expressed as,
\begin{equation}
    \tilde{b_{ij}}(\tau)=R_{u_{i}u_{j}}(\tau), \ i,j=1,2,3.
\end{equation}
The diagonal components of $\tilde{b_{ij}}(\tau)$ represent $R_{xx}(\tau)$ (where $x$ can be $u,v,w$) values, which quantify the intermittency effects on the velocity variances at each scale of the flow. On the other hand, the cross-diagonal terms are associated with large intermittent fluctuations in the instantaneous flux components. For instance, a particular cross-diagonal component of $\tilde{b_{ij}}(\tau)$, represents,
\begin{equation}
    R_{uw}(\tau)=\frac{\mathcal{B}(\Delta u \Delta w)}{\mathcal{B}(\Delta u_{\rm p} \Delta w_{\rm p})}-1,
\end{equation}
where $\mathcal{B}(\Delta u \Delta w)$ denote the burstiness indices of the instantaneous streamwise momentum flux at a scale $\tau$ \citep{chowdhuri2023revisiting}, while the subscript $\rm p$ denote the momentum flux signals obtained from the IAAFT model of the two velocity components. Physically, as this cross component approaches zero, it implies that the flux generation at that particular scale is not sensitive to the presence of small-scale intermittency. If the three eigen values of $\tilde{b_{ij}}$ are $|\tilde{\lambda_{i}}|$, where $i=1,2,3$, then $\tilde{b_{ij}}$ can be diagonalized as,
\begin{equation}
\tilde{b_{ij}}=
  \begin{bmatrix}
    |\tilde{\lambda_{1}}| & 0 & 0 \\ \\
    0 & |\tilde{\lambda_{2}}| & 0 \\ \\
    0 & 0 & |\tilde{\lambda_{3}}|
  \end{bmatrix}
  \end{equation}
In case of an isotropic configuration, one would expect to satisfy the condition of $|\tilde{\lambda_{1}}|=|\tilde{\lambda_{2}}|=|\tilde{\lambda_{3}}|$, thereby indicating that no directional preference exists in how the small- scale intermittency affects the three different velocity components. Note that we use the absolutes of the eigen values since the magnitudes of these quantities matter the most rather than their signs \citep{betchov1956inequality}. As shown in Appendix \ref{app_A}, out of these three eigen values, $|\tilde{\lambda_{3}}|$ is the largest, and their scale-wise variations remain remarkably consistent among the DNS, ASL, and RSL datasets. After studying the behaviour of the eigenvalues in Fig. \ref{fig:6}, the anisotropy in $\tilde{b_{ij}}$ can be conveniently expressed through the metric,
\begin{equation}
\tilde{\lambda_{\rm eff}}=\frac{|\tilde{\lambda_{3}}|}{\sqrt{\frac{{\tilde{\lambda}}^2_{1}+{\tilde{\lambda}}^2_{2}}{2}}}
\end{equation}
whose values approach unity as the differences in the three eigen values decrease and are significantly larger than 1 when anisotropy persists. This formulation is qualitatively similar to \citet{pumir2016small}, where a somewhat similar metric was used to study the anisotropy in the velocity strain tensor. 

\begin{figure*}[h]
\centerline{\includegraphics[width=\textwidth]{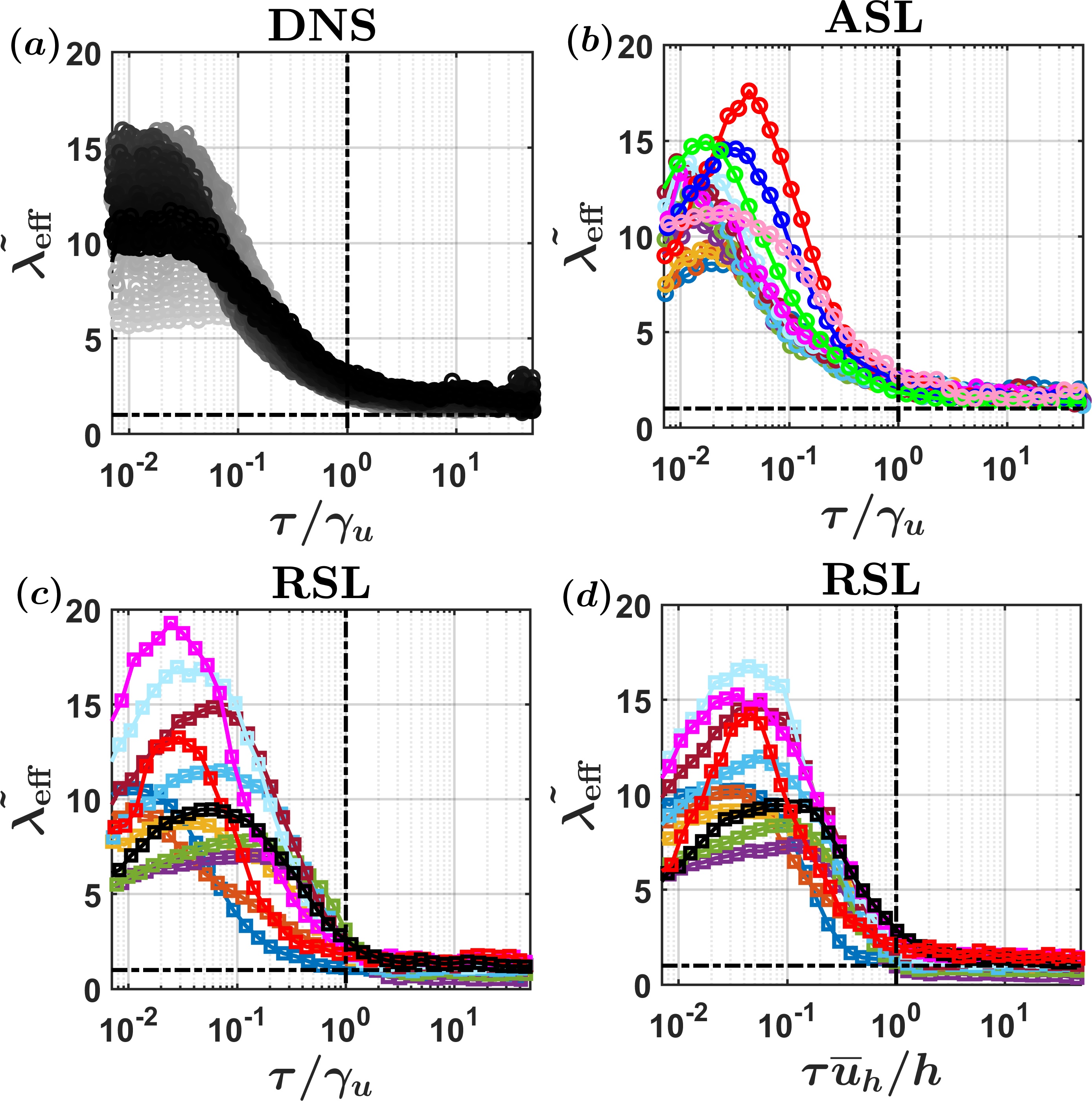}}
\caption{The effective eigenvalues ($\tilde{\lambda_{\rm eff}}$) of the intermittent Reynolds stress tensor are shown for the (a) DNS, (b) ASL, and (c) RSL datasets. (d) Same as in (c), but the timescales are normalized by the canopy shear time scale $\overline{u}_{h}/h$, where $\overline{u}_{h}$ is the mean wind-speed at $z=h$ level. The legends are same as in Fig. \ref{fig:2}.} 
\label{fig:4}
\end{figure*}

In Fig. \ref{fig:4} we show how $\tilde{\lambda_{\rm eff}}$ varies with $\tau/\gamma_u$ across all the three different datasets. The three different panels in Fig. \ref{fig:4} correspond to the DNS (Fig. \ref{fig:4}a), ASL (Fig. \ref{fig:4}b), and RSL (Fig. \ref{fig:4}c) observations, respectively. It is clear that irrespective of the dataset types, $\tilde{\lambda_{\rm eff}}$ values decrease with increasing scales and at scales larger or equal to $\gamma_u$ they approach unity. Therefore, a significant amount of anisotropy persists at smaller scales, which systematically disappears as the larger scales are encountered. In Appendix \ref{app_B}, a thorough comparison is presented among the three datasets (see Fig. \ref{fig:7}), and it is apparent that the $\tilde{\lambda_{\rm eff}}$ values from the logarithmic layers of DNS and ASL datasets agree quite well with each other. On the other hand, the RSL datasets show a clear difference with the ASL ones. Specifically, the RSL datasets display a clear peak in the $\tilde{\lambda_{\rm eff}}$ curves, which indicates that there is a particular scale where the anisotropy is the largest for such flows. These peak positions do not collapse exactly when $\tau$ are scaled with $\gamma_u$. In fact, a nice collapse is observed when the canopy shear scale is used (see Fig. \ref{fig:4}d). This collapsed peak position is at $\tau \overline{u}_{h}/h \approx 0.04$, which further underscores the importance of the canopy-scale coherent structures in determining the features of small-scale anisotropy. However, the DNS and ASL datasets too display a peak in their $\tilde{\lambda_{\rm eff}}$ curves and the location of these peaks collapse reasonably well at a scale of $\tau/\gamma_{u} \approx 0.03-0.04$. 

Nevertheless, upon close inspection, it is apparent that the magnitudes of $\tilde{\lambda_{\rm eff}}$ values indicate a clear height dependency for the RSL datasets, such that their values remain the largest at heights above the canopy. This height dependency is not very prominent for the ASL datasets. It is, therefore, interesting to ask why such differences exist and whether there exists any connection with the anisotropy of the velocity field at energy-containing scales (i.e., comparable to the integral scales of $u^{\prime}$) of the flow.

\subsection{Small- and Large-scale anisotropy}
\label{Res3}
\begin{figure*}
\centerline{\includegraphics[width=\textwidth]{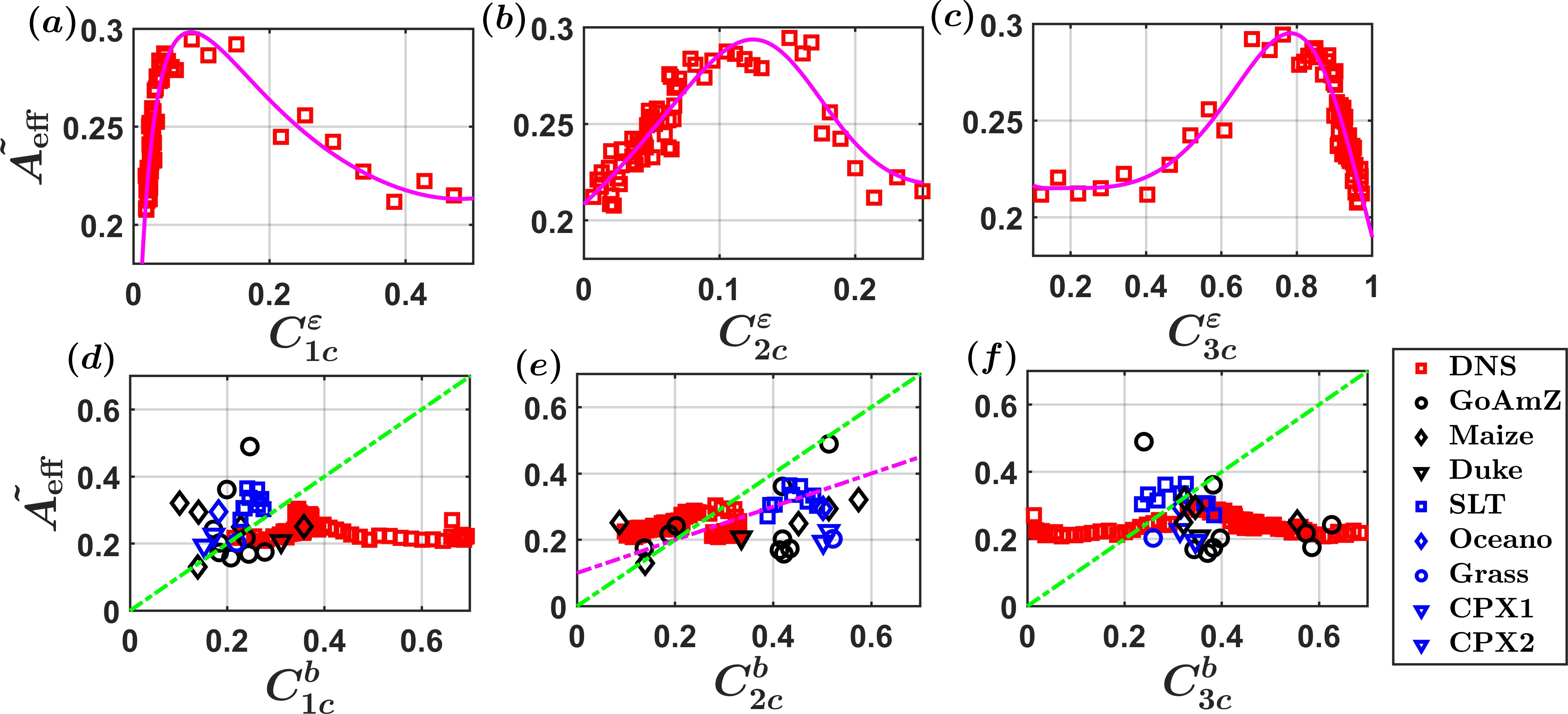}}
\caption{A bulk measure of small-scale anisotropy ($\tilde{A_{\rm eff}}$), obtained by integrating the $\tilde{\lambda_{\rm eff}}$ values between 0 and $\gamma_{u}$, is compared between the three anisotropic states of the (a)--(c) dissipation ($C^{\varepsilon}_{i\rm c}$, where $i=1,2,3$) and (d)--(f) Reynolds stress tensors ($C^{b}_{i\rm c}$, where $i=1,2,3$). The red colored squares denote the DNS dataset, while the blue and black coloured markers represent the ASL and RSL datasets. The thick pink lines in (a)--(c) indicate a non-linear fit to the data and in (d)--(f) the green dash-dotted lines indicate the 1:1 straight line. The pink dash-dotted line in (e) denote a line with a slope of $0.5$.} 
\label{fig:5}
\end{figure*}

To create a bulk measure of small-scale anisotropy, we integrate the $\tilde{\lambda_{\rm eff}}$ values up to the scales $\tau=\gamma_{u}$. If no anisotropy was present within those scales, this area would be exactly equal to 1. Hence, a bulk measure of small-scale anisotropy can be defined as,
\begin{equation}
\tilde{A_{\rm eff}}=\frac{1}{\int_{0}^{\gamma_{u}} {\tilde{\lambda_{\rm eff}}(\tau) \,d\tau}}.
\end{equation}
Since $\tilde{\lambda_{\rm eff}}$ values are always greater than unity when anisotropy persists (see Fig. \ref{fig:4}), $\tilde{A_{\rm eff}}$ values are bounded between $0 \leq \tilde{A_{\rm eff}} \leq 1$. For practical purposes, to numerically compute the integral, ${\int_{0}^{\gamma_{u}} {\tilde{\lambda_{\rm eff}}(\tau) \,d\tau}}$, we use a trapezoidal approximation. 

In general, the small-scale anisotropy is studied through the dissipation tensor \citep{antonia1994anisotropy,longo2017invariants}, whose values can only be computed from the DNS data. Such computation is not possible from the atmospheric data, since the dissipation tensor requires measurements of the whole three-dimensional flow field. On the contrary, the estimation of $\tilde{A_{\rm eff}}$ can be accomplished from any dataset due to the well-behaved nature of $\tilde{\lambda_{\rm eff}}$ curves (see Fig. \ref{fig:4}), which can be computed for any single-point turbulence measurements. 

As a first order check, we compare $\tilde{A_{\rm eff}}$ values with the anisotropic states of the dissipation tensor ($d_{ij}$, see Eq. \ref{tensor}), estimated from our DNS dataset. The partial derivatives of the velocity components, as required for the dissipation tensor, are calculated from a forward-differencing scheme such that, $\partial{u}/\partial{x}=(u^{\prime}[i+1,j,k]-u^{\prime}[i,j,k])/\Delta x$, where $i,j,k$ are the grid-point coordinates of the DNS data at $x,y,z$ directions, respectively. The anisotropic states of the dissipation tensor are defined from the perspective of a barycentric map \citep{banerjee2007presentation}, whose three components are $C^{\varepsilon}_{1\rm c}$, $C^{\varepsilon}_{2\rm c}$, and $C^{\varepsilon}_{3\rm c}$. These three components are determined as,
\begin{align}
\begin{split}
C^{\varepsilon}_{1 \rm c} &= \lambda_{1}-\lambda_{2}
\\
C^{\varepsilon}_{2 \rm c} &= 2(\lambda_{2}-\lambda_{3})
\\
C^{\varepsilon}_{3 \rm c} &= 3\lambda_{3}+1
\end{split},
\label{bm}
\end{align}
where $\lambda_{1}$, $\lambda_{2}$, $\lambda_{3}$ are the three eigenvalues of the dissipation tensor in the order $\lambda_{1} > \lambda_{2} > \lambda_{3}$. The comparisons between $\tilde{A_{\rm eff}}$ and $C^{\varepsilon}_{i \rm c}$ (where $i=1,2,3$) are shown in Figs. \ref{fig:5}a--c. It is clear from the figures that these two metrics are definitely related to each other but through a non-linear relationship, expressed as,
\begin{equation}
\tilde{A_{\rm eff}}=\frac{P(C^{\varepsilon}_{i \rm c})}{Q(C^{\varepsilon}_{i \rm c})},  
\end{equation}
where $P$ and $Q$ are the fifth-order polynomial expansions of $C^{\varepsilon}_{i \rm c}$, whose coefficients are estimated from a data fitting exercise. One point to note is the ranges for $\tilde{A_{\rm eff}}$ and $C^{\varepsilon}_{i \rm c}$ are different. This difference could have occurred since the dissipation tensor is an area-averaged measure while the $\tilde{A_{\rm eff}}$ values are obtained by integrating $\tilde{\lambda_{\rm eff}}(\tau)$ up to a certain scale, which is $\gamma_{u}$.

At the same time, $\tilde{A_{\rm eff}}$ values are also compared with the three anisotropic states ($C^{b}_{i \rm c}$) of the Reynolds stress tensor $b_{ij}$ (see Eq. \ref{reynolds}), defined analogously as in Eq. \ref{bm}. These anisotropic states can be computed for all the datasets and from Figs. \ref{fig:5}d--f, it is apparent that the $\tilde{A_{\rm eff}}$ values are very strongly connected to the two-component anisotropic state of Reynolds stress tensor ($C^{b}_{2\rm c}$) rather than to the one ($C^{b}_{1\rm c}$)- or three-($C^{b}_{3\rm c}$) component states. To provide a sense of reference, the green lines in Fig. \ref{fig:5}d--f indicate a straight line of $45^{\circ}$ slope, from which it is evident that the data-points are quite scattered around it without any underlying order when $C^{b}_{1\rm c}$ or $C^{b}_{3\rm c}$ are considered. Contrary to this, the strong relationship between $\tilde{A_{\rm eff}}$ and $C^{b}_{2\rm c}$ holds irrespective of the Reynolds number or surface conditions of the flow, thereby hinting towards a universal behaviour. Physically, the two-component anisotropic state of the Reynolds stress tensor indicates the influence of the horizontal motions over the vertical ones on the turbulence statistics. From a topological perspective, the two-component anisotropy is connected to the presence of large-scale coherent structures near the wall whose vertical components are blocked due to the location of the wall itself \citep{chowdhuri2020revisiting}.

From a practical standpoint, the relationship between $\tilde{A_{\rm eff}}$ and $C^{b}_{2 \rm c}$ can be well-represented through a straight line of a slope of 0.5, or in other words, $\tilde{A_{\rm eff}}=C^{b}_{2 \rm c}/2$. The $R^{2}$ value associated with this fit is larger than 0.9 and thus can be considered to be statistically robust. By knowing such relationship, it is possible to infer about the presence of small-scale anisotropy in the flow from the large-scale state itself. As shown by \citet{antonia1994anisotropy}, the information about small-scale anisotropy is quite important to refine the standard $k$-$\varepsilon$ model of wall turbulence, where it is implicitly assumed that at smaller scales the turbulence tend to be isotropic. As a future work, it remains to be seen whether our empirical relationship can be directly used to separate the dissipation rate to an isotropic and an anisotropic part.

\section{Conclusion}
\label{conclusion}
In this study a scale-wise analysis of small-scale intermittency is introduced and applied over a range of numerical and experimental datasets with the Reynolds numbers varying from $10^3$ to $10^6$, the surface conditions spanning from aerodynamically smooth surfaces to grasslands to forests having trees as large as 35 m, and the flow types being considered encompass channel flows to boundary layers to atmospheric surface layers and roughness sublayers. For such a wide variety of datasets, our findings indicate that the effects of small-scale intermittency persists up to the scales of the order of integral scales of the streamwise velocity fluctuations. Therefore, this scale provides a universal basis to separate the effects of large-scale flow features from the small-scale ones.

Moreover, we find that the effects of intermittency is very different for the horizontal and vertical velocity components, in terms of their scaling properties and magnitudes. This conclusion also appears to be universal, and therefore, we use this information further to define a metric for small-scale anisotropy. This metric is based on the eigen values of an intermittent Reynolds stress tensor, whose properties are quite sensitive to the presence of small-scale flow features. Unlike the dissipation tensor, whose computation is only limited to three-dimensional numerical datasets, our metric can be easily computed for any point-wise experimental measurements, whether obtained from engineering or atmospheric flows.

We show that the effect of small-scale anisotropy is mainly determined by the presence of coherent structures in the flow. Based on this finding, a diagnostic relationship is proposed between the small-scale anisotropy and the two-component anisotropic state of the Reynolds stress tensor. This relationship remarkably holds over a wide range of flows and for practical purposes, can be used to refine the $k$-$\varepsilon$ models of wall turbulence where the assumption of isotropy at smaller scales plays a pivotal role. In conclusion, we address an important gap in turbulence research, concerned with the fact that whether the small-scale turbulence features are universal or not. It turns out there are indeed a couple of aspects that can be considered universal. First, the scales up to which the small-scale effects continue, and second, the relationship between small-scale anisotropy and large-scale coherent structures. 

As a possible limitation, this study is confined to neutral conditions, and therefore, in future, it would be interesting to investigate the role of buoyancy on (a) small-scale intermittency; (b) the scaling properties of $R_{uu}$ and $R_{ww}$ curves; and (c) their anisotropic characteristics. One another future direction is scalar turbulence, where by studying the $R_{xx}$ (where $x$ could be temperature, carbon-dioxide, or water-vapor fluctuations) curves, the topic of intermittency and scalar similarity at the smaller scales of the flow can be addressed for both convective and neutral stratification.

\section*{Acknowledgements}
SC and TB acknowledge the funding support from the US National Science Foundation ({\it NSF-AGS-PDM-2146520}, {\it NSF-OISE-2114740}, {\it NSF-CPS-2209695}, \textit{NSF-ECO-CBET-2318718}, and \textit{NSF-DMS-2335847}), the University of California Office of the President (UCOP-LFR-20-653572),  NASA (\textit{80NSSC22K1911}), and the United States Department of Agriculture ({\it NIFA  2021-67022-35908}, and \textit{USDA-20-CR-11242306-072}). CAIPEEX-IGOC experiment was funded by the Ministry of Earth Sciences, Govt of India and SC acknowledges the entire CAIPEEX team and specially Thara Prabhakaran and Anand Karipot for kindly letting him use the tower data for this research. SC thanks Giovanni Iacobello for some initial help and discussion regarding the processing of the DNS dataset and sharing the code to read the data. The authors acknowledge Khaled Ghannam, Marcelo Chamecki, Jasper Kok, and Gaby Katul for the use of the GoAmazon, Maize, Oceano, DF, and Grass datasets. The codes developed in this study can be shared with the interested researchers by contacting the corresponding author.

\appendix
\section{The eigen values of $\tilde{b_{ij}}$}
\label{app_A}
\begin{figure*}[h]
\centerline{\includegraphics[width=\textwidth]{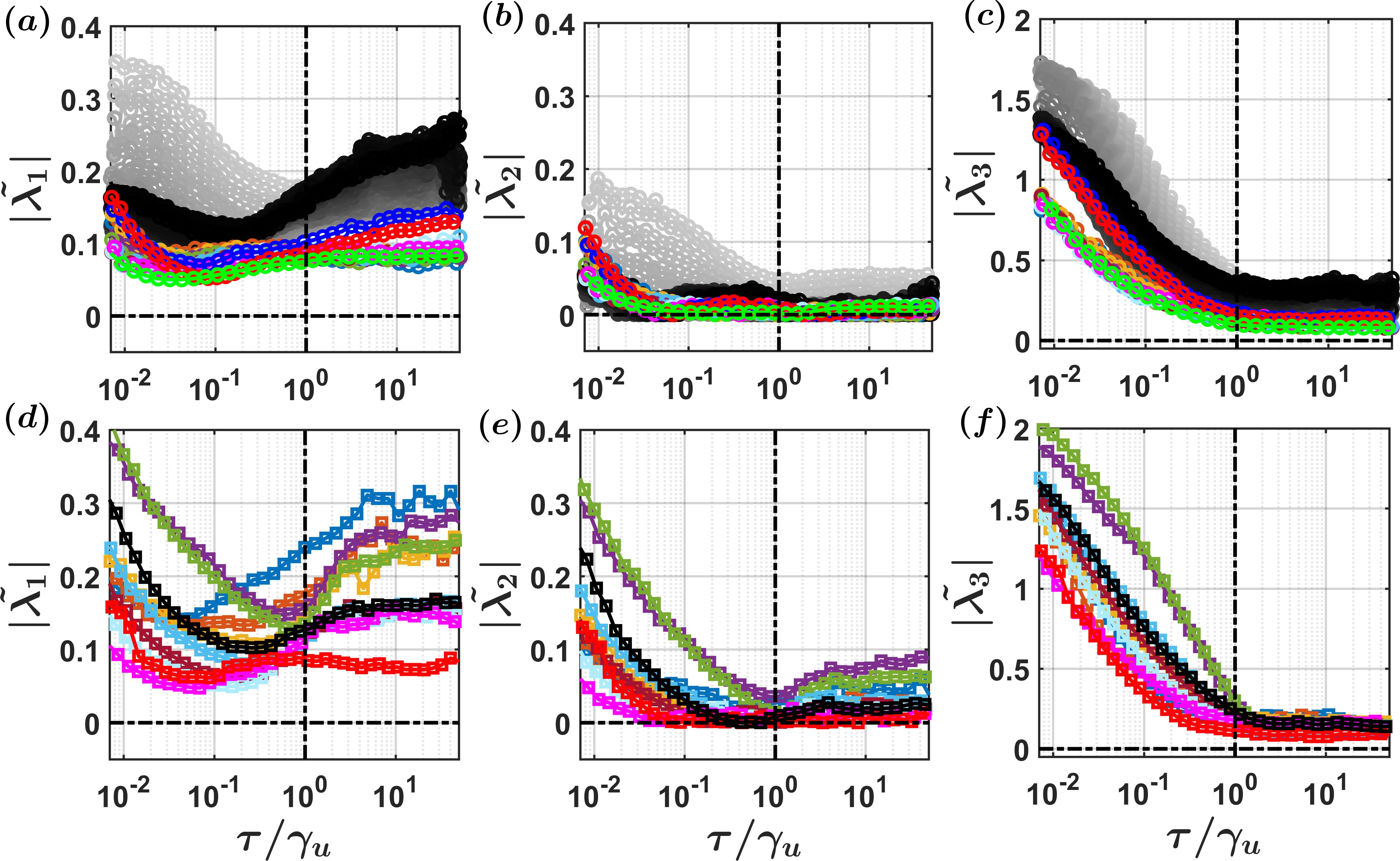}}
\caption{The magnitudes of the three eigen values ($|\tilde{\lambda_{i}}|$, where $i=1,2,3$) of the intermittent Reynolds stress tensor are shown. The upper three panels (a)--(c) indicate the plots for the DNS and ASL datasets, while the lower three panels (d)--(f) indicate those from the RSL flows. The colors represent the same information as in Fig. \ref{fig:2}.} 
\label{fig:6}
\end{figure*}
In this appendix we show the behaviour of the three eigen values ($|\tilde{\lambda_{i}}|$, where $i=1,2,3$) of the intermittent Reynolds stress tensor with $\tau/\gamma_u$. We take the magnitudes of the eigen values instead of their original signs, which is considered as a standard practice in the community \citep{betchov1956inequality,suzuki2003absolute,tanahashi2004scaling}. The upper three panels of Fig. \ref{fig:6} compare the eigenvalues between the DNS and ASL datasets (Fig. \ref{fig:6}a--c), while the lower three panels show the same for the RSL datasets (Fig. \ref{fig:6}d--f). It is apparent that, irrespective of the datasets considered, the qualitative behaviour of the three eigen values with $\tau/\gamma_u$ remain very similar among all the three different datasets. For instance, the eigenvalues $|\tilde{\lambda_{3}}|$ remain the largest of the three and decrease monotonically with increasing scales. This behaviour is very similar to how the individual $R_{uu}$ or $R_{ww}$ curves behave (see Figs. \ref{fig:2} and \ref{fig:3}), and therefore, the effects of small-scale intermittency can be fully described by $|\tilde{\lambda_{3}}|$ alone. In fact, similar to $R_{uu}$ curves, for RSL datasets, the values of $|\tilde{\lambda_{3}}|$ at scales $\tau<\gamma_u$ remain the strongest for those GoAmazon levels where the leaf area densities are the largest. On the other hand, the eigenvalues $|\tilde{\lambda_{1}}|$ and $|\tilde{\lambda_{2}}|$ (with $|\tilde{\lambda_{2}}|$ being the smallest) are considerably smaller than $|\tilde{\lambda_{3}}|$ for scales $\tau<\gamma_u$, thereby indicating the presence of anisotropy at smaller scales with the effect of intermittency being different for the velocity field between the horizontal and vertical directions. It is interesting to note that, contrary to the other two eigenvalues, $|\tilde{\lambda_{1}}|$ values do not monotonically decrease with increasing scales, rather they attain a minimum at a particular scale beyond which they increase again.

\section{Comparison of $\tilde{\lambda_{\rm eff}}$ values}
\label{app_B}
\begin{figure*}[h]
\centerline{\includegraphics[width=\textwidth]{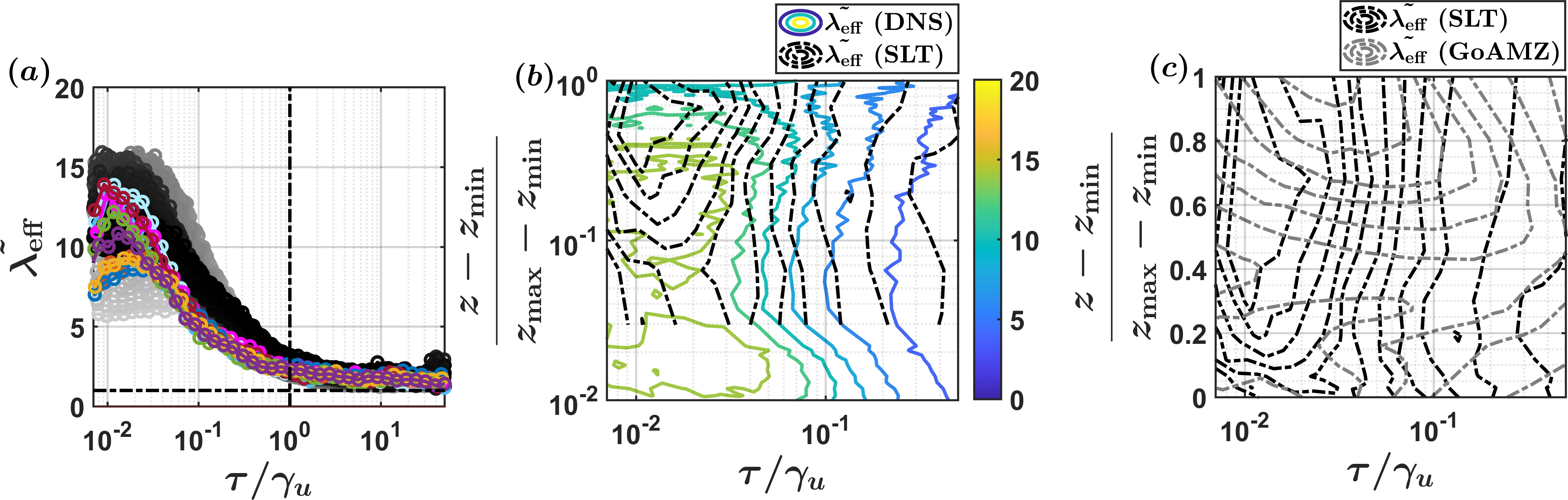}}
\caption{(a) The $\tilde{\lambda_{\rm eff}}$ values are compared between the DNS (grey lines) and SLT (colored lines) datasets. (b) The same as (a), but shown as a contour plot where the heights are normalized as $(z-z_{\rm min})/(z_{\rm max}-z_{\rm min})$. The black contours represent $\tilde{\lambda_{\rm eff}}$ values from the SLT dataset while the colored ones are from the DNS. (c) The contours of $\tilde{\lambda_{\rm eff}}$ values are compared between the SLT (black contour lines) and GoAMZ (grey contour lines) datasets.} 
\label{fig:7}
\end{figure*}
A detailed comparison of $\tilde{\lambda_{\rm eff}}$ values is presented in Fig. \ref{fig:7}, among the DNS, ASL, and RSL datasets. For clarity purposes, regarding the ASL and RSL flows, the SLTEST and GoAmazon datasets are considered to be the representative ones, since they contain measurements from multiple levels. In Fig. \ref{fig:7}a, we directly overlay the $\tilde{\lambda_{\rm eff}}$ values obtained from the SLTEST dataset on the DNS ones. From a visual inspection, it appears that the SLTEST curves qualitatively follow the DNS ones quite well as the logarithmic layer is approached. The lines with darker shades from the DNS datasets indicate the logarithmic layer. However, to quantify this aspect more precisely, in Fig. \ref{fig:7}b, we construct a contour plot, where the contour values represent the variations in $\tilde{\lambda_{\rm eff}}$. To differentiate between the two datasets, the colored contours of $\tilde{\lambda_{\rm eff}}$ values indicate the DNS dataset while the black ones are from the SLTEST data. Moreover, since the height ranges of the two datasets are significantly different, we employ a standard normalization where the heights ($z$) are scaled as, $(z-z_{\rm min})/(z_{\rm max}-z_{\rm min})$. After applying this scaling, it becomes clear from Fig. \ref{fig:7}b that the SLTEST contours match the DNS ones quite well as the heights of the DNS dataset increase. Therefore, despite huge differences in their Reynolds numbers nearly by three orders of magnitude, a similarity emerges between the channel and ASL flows. On the other hand, when the contours of $\tilde{\lambda_{\rm eff}}$ values are compared between the SLTEST and GoAmazon datasets, significant differences appear between the two (see Fig. \ref{fig:7}c). This highlights the influence of the roughness elements on the small-scale turbulence statistics, which can be physically accounted for through a canopy shear scale as pointed out in Fig. \ref{fig:4}.

\bibliography{References}% Produces the bibliography via BibTeX.

\end{document}